\documentclass[12pt,preprint]{aastex}
\shorttitle{Gamma-rays from GC MSPs}
\shortauthors{Venter & de Jager}

\begin{document}
\title{Predictions of Gamma-ray Emission from Globular Cluster Millisecond Pulsars Above 100~MeV}

\author{C. VENTER,\altaffilmark{1,2,3} O.C. DE JAGER,\altaffilmark{2,3,4} AND A.-C. CLAPSON\altaffilmark{5}}
%\email{Christo.Venter@nwu.ac.za}
\altaffiltext{1}{NASA Postdoctoral Program Fellow, Astrophysics Science Division, NASA/Goddard Space Flight Center, Greenbelt, MD 20771}
\altaffiltext{2}{Unit for Space Physics, North-West University, Potchefstroom Campus, Private Bag X6001, Potchefstroom 2520, South Africa}
\altaffiltext{3}{Centre for High Performance Computing, CSIR Campus, 15 Lower Hope Street, Rosebank, Cape Town, South Africa}
\altaffiltext{4}{South African Department of Science and Technology, and National Research Foundation Research Chair: Astrophysics and Space Science}
\altaffiltext{5}{Max-Planck-Institut fuer Kernphysik, P.O.\ Box 103938, D~69029 Heidelberg, Germany}

\begin{abstract}
The recent \textit{Fermi} detection of the globular cluster (GC) 47~Tucanae highlighted the importance of modeling collective gamma-ray emission of millisecond pulsars (MSPs) in GCs. Steady flux from such populations is also expected in the very high energy (VHE) domain covered by ground-based Cherenkov telescopes. We present pulsed curvature radiation (CR) as well as unpulsed inverse Compton (IC) calculations for an ensemble of MSPs in the GCs 47~Tucanae and Terzan~5. We demonstrate that the CR from these GCs should be easily detectable for \textit{Fermi}, while constraints on the total number of MSPs and the nebular B-field may be derived using the IC flux components.
\end{abstract}


\section{Introduction}
\label{sec:intro}
The launch of \textit{Fermi Gamma-ray Space Telescope (Fermi)} on 11~June~2008 heralded a new era for high-energy (HE) Gamma-ray Astronomy. The \textit{Large Area Telescope (LAT)} aboard \textit{Fermi} \citep{Atwood09} is $\sim25$ times more sensitive than its predecessor, the \textit{Energetic Gamma Ray Experiment Telescope (EGRET)}. 
%and detected the six known \textit{EGRET} pulsars within its 60-day early operations period \citep{Parkinson09}. 
Recently, \textit{Fermi} anounced the detection of the globular cluster (GC) 47~Tucanae as a bright point source \citep{Abdo09}. This emission may plausibly be due to the large number of millisecond pulsars (MSPs) residing in this cluster \citep[e.g.][hereafter HUM05]{HUM05}.

GCs are furthermore possible sources of very high energy (VHE) gamma rays \citep[][hereafter BS07]{BS07}, and are / will be important targets for ground-based Cherenkov telescopes such as the \textit{High Energy Stereoscopic System} \citep[\textit{H.E.S.S.} --][]{Aharonian06} as well as future telescopes such as the \textit{Cherenkov Telescope Array (CTA)}\footnote{www.cta-observatory.org} and the \textit{Advanced Gamma-ray Imaging System (AGIS)}\footnote{http://gamma1.astro.ucla.edu/agis/index.php/Main\_Page}.

\citet{VdeJ08a} modeled the collective curvature radiation (CR) expected from 47~Tucanae, demonstrating that this pulsed component should be detected by \textit{Fermi}, depending on the number of visible gamma-ray pulsars in the GC MSP population, $N_{\rm vis}$ (see \S\ref{sec:intro2}). \citet{VdeJ08b} extended these results by modeling the unpulsed inverse Compton (IC) and synchrotron radiation (SR) flux components expected due to the interaction of HE electrons, ejected from MSP magnetospheres, with the ambient cosmic microwave background (CMB) and bright starlight photons, in the presence of a nebular magnetic field $B$.

In this Letter, we give updated CR spectra (\S\ref{sec:pulsed}, see \citet{VdeJ09} for details), in addition to further IC calculations (\S\ref{sec:unpulsed}), for 47~Tucanae and also Terzan~5, another GC containing many MSPs (\S\ref{sec:intro2}). The roles of \textit{Fermi, H.E.S.S.}, and \textit{CTA} for deriving model constraints are discussed in \S\ref{sec:con}.

\section{Globular Cluster Millisecond Pulsars}
\label{sec:intro2}
GCs are very old galactic sub-structures, with evolved stellar composition. As a consequence, they are likely to harbor a larger than usual density of end products of stellar evolution -- compact objects. This is supported by observations of X-ray binary systems and MSP populations in GCs. Since the discovery of the first GC MSP in M28 \citep{L87}, a total of~140 pulsars have been discovered in~26 GCs \citep{F09}, with 90\% having periods $P < 20$~ms \citep{Ransom08}. 

Among the GCs, 47~Tucanae and Terzan~5 stand out by the number of radio-detected MSPs they host: 23 and 33 respectively \citep{Ransom08}. However, these GCs are otherwise remarkable for different reasons. 47~Tucanae is one of the most massive GCs, after $\omega$~Centauri, while the core density of Terzan~5 gives it the highest expected rate of binary interactions \citep{Pooley06}. Basic properties for these two objects are summarized in Table~\ref{tab:andre}.

Besides pulsed emission, detected already in radio, X-rays, and gamma rays, MSPs are also expected to produce steady low-flux VHE emission. Given the sensitivity of current instruments, it is likely that only a population of such objects would produce detectable VHE fluxes (\S\ref{sec:unpulsed}). The extension of the cores of the GCs is furthermore clearly below the angular resolution of current HE and VHE instruments. In this Letter we are therefore not concerned about individual MSP fluxes, but attempt to constrain a particular class of polar cap (PC) pulsar models using the total flux from a GC MSP ensemble. This approach reduces uncertainties in the CR and IC fluxes due to unknown pulsar geometries, as the relative uncertainties in ensemble-averaged fluxes, which scale as the reciprocal of the square root of the number of MSPs \citep{VdeJ08a}, are typically one order of magnitude smaller (for $\sim100$~MSPs) than for the case of single MSP `geometry-averaged' fluxes.

An important issue in the context of this work is the uncertainty on the size of the MSP population in each system, beyond the radio-detected ones. We dinstinguish between two numbers. We define the number of visible \textit{gamma-ray} pulsars ($N_{\rm vis}$) as those MSPs whose CR is (mostly) beamed toward earth, so that observers will in principle be able to measure pulsed emission from these (although it might not always be possible to distinguish individual members due to experimental limitations). In contrast, $N_{\rm tot}$ represents the total number of MSPs in the GC, whether visible in pulsed gamma rays or not. 

Mostly off-axis pulsars might still contribute to the total CR flux, as observers may be clipping the faint edges of gamma-ray beams from these pulsars. We therefore include them in $N_{\rm vis}$, along with the on-beam MSPs. HUM05 noted that gamma-ray beams of MSPs are probably wider than their radio beams, so $N_{\rm vis}$ is probably larger when applied to gamma-ray pulsars (as done here) than when applied to radio ones, possibly up to $N_{\rm vis} \simeq N_{\rm tot}$. The IC flux however depends on $N_{\rm tot}$, as \textit{all} pulsars, including ones with off-beam geometries, are expected to contribute VHE leptons (mostly electrons, as most MSPs are believed to be pair production starved; see HUM05) which upscatter soft photons to gamma-ray energies. 

The isolated location of 47~Tucanae in the Southern sky away from the Galactic Plane made it a prime target for GC studies. In radio, \citet{McConnell04} estimated from the unresolved emission a maximum of 30 detectable MSPs. In X-ray, \textit{Chandra} observations revealed a large number of unidentified sources in the core of 47~Tucanae \citep{H05}, from which followed an upper limit of about 60 MSPs, based on the X-ray properties of radio-confirmed MSPs in the GC.

For Terzan~5, the available observations are not so constraining. Again in the radio, \citet{Fruchter00} estimated a range of $60 - 200$ MSPs in the core of the GC, from \textit{Very Large Array (VLA)} and \textit{Australia Telescope Compact Array (ATCA)} data. No equivalent result has yet been derived in X-rays. The analysis of \textit{Chandra} data by \citet{Heinke06} yielded X-ray sources, but with too high a sensitivity limit to constrain the MSP population.

Finally, for both systems, GC evolution models \citep[e.g.][]{Ivanova08} suggest MSP numbers below~200, of which a fraction would be detectable. 

\begin{deluxetable}{lcccc} 
\tablecolumns{4} 
\tablewidth{0pc} 
\tablecaption{Selected Globular Cluster Characteristics\label{tab:andre}} 
\tablehead{ 
\colhead{} & \colhead{47~Tucanae}   & \colhead{Reference}    & \colhead{Terzan~5} & 
\colhead{Reference}}
\startdata 
Distance $d$ (kpc) & 4.0 & 1 & 5.5 $-$ 8.7 & 2, 3\\
Mass ($M_\odot$) & $1.0\times10^6$ & 4 & $3.5\times10^5$ & 5 \\
Core radius $r_{\rm c}$ (arcmin) & 0.40 & 6 & 0.18 & 6\\
Half mass radius $r_{\rm hm}$ (arcmin) & 2.79 & 6 & 0.83 & 6\\
Detected radio MSPs & 23 & 7 & 33 & 7 \\
Total stellar luminosity ($L_\odot$) & $7.5\times10^5$ & 8 & $1.5\times10^5$ & 8
\enddata 
\tablerefs{
(1) \citet{McLaughlin06};
(2) \citet{Ortolani07};
(3) \citet{Cohn02};
(4) \citet{McLaughlin05};
(5) \citet{Ivanova08};
(6) \citet{C09}, extended from \citet{Harris96};
(7) \citet{F09};
(8) BS07.}
\end{deluxetable}

\section{Pulsed Gamma-ray Flux}
\label{sec:pulsed}
As previously \citep{VdeJ08a}, we use an isolated pulsar PC model \citep{VdeJ05}, including the effect of General Relativistic (GR) frame dragging \citep[e.g.][]{MH97, HM98}. In accordance with the expectation that most MSPs are inefficiently screened by CR and IC scattering pairs (HUM05), we use an unscreened acceleration potential. We furthermore use the same population of MSPs: $\dot{P}$-values (period-derivatives) from \citet{B06}, which imply an average spin-down luminosity of $\dot{E}_{\rm rot}=2.2\times10^{34}$~erg\,s$^{-1}$, as well as canonical values for equation of state (EOS) parameters (moment of inertia $I_{\rm NS}=0.4MR^2$, stellar radius $R=10^6$~cm, and pulsar mass $M=1.4M_\odot$). As previously, we add the CR spectra of $N_{\rm vis}=100$ MSPs, randomly selected from our population of 13 members (each with a random pulsar geometry, i.e.\ magnetic inclination and observer angle, including off-beam cases) iteratively a million times, and so obtain an average pulsed spectrum, with 2$\sigma$-bands indicating the uncertainty due to varying $P$ and $\dot{P}$-values, as well as the unknown pulsar geometries. 

\citet{VdeJ08a} used a delta function approximation for the CR spectrum radiated per primary, but the resulting spectral shape did not correspond well to results of similar studies \citep[e.g.][]{Frackowiak05}, as pointed out by Dyks (2008, personal communication). We therefore developed a corrective technique involving the full CR spectrum per primary \citep{VdeJ09}, and applied it here.
%The mismatch between our old and new CR spectra stems from the fact that we previously \citep{VdeJ08a} used a delta function approximation for the total CR spectrum from each electron at a given position along the MSP B-field, instead of assigning the full spectrum of the form $(dN/dE)_{\rm CR}\propto E^{-2/3}\exp(-E/E_0)$, with $E_0$ the CR cut-off energy. The total power in both the old and updated spectra is the same, but the high and low-energy tails of the final spectra were underestimated using the old approach. The corrected single particle CR spectra are added to give the total CR spectrum from each MSP as a function of pulsar geometry. In turn, addition of MSP CR spectra gives the average cluster CR spectrum.

Assuming that the MSPs in Terzan~5 have similar basic pulsar parameters than those in 47~Tucanae, we scaled the cumulative CR spectrum of the latter to distances of $d=5.5$~kpc and 8.7~kpc. The resulting updated CR spectra for 47~Tucanae and Terzan~5 are shown in panels~(a) and~(b) of Figure~\ref{fig:Diff}. 

%It is probably important to reiterate the point made by \citet{VdeJ08a} that a population of MSPs has lower relative ensemble errors on the average spectrum than in the case of single MSPs, so that derivision of much firmer constraints on average pulsar parameters is hereby facilitated.

\section{Unpulsed Gamma-ray Flux}
\label{sec:unpulsed}
We calculate the \textit{ejection} spectrum of electrons leaving each MSP by binning the number of primary electrons ejected per unit time according to their residual energies at the \textit{light cylinder}. %Similar to \S\ref{sec:pulsed}, 
We add the spectra of $N_{\rm tot}=100$~MSPs (with random inclination angles) and iterate the Monte Carlo procedure a million times. From this we obtain an average cumulative ejection particle spectrum and 2$\sigma$-bands \citep{Venter_phd}. 

We divide the region where unpulsed radiation is generated into two zones: $Z0$, reaching from $r=0$ to $r=r_{\rm c}$, with $r_{\rm c}$ the core radius, and $Z1$, reaching from $r=r_{\rm c}$ to $r=r_{\rm hm}$, with $r_{\rm hm}$ the half mass radius (see Table~\ref{tab:andre}). For 47~Tucanae, we used energy densities of $u^0_{\rm rad}\sim3\,000$ eV/cm$^3$ for $Z0$, and $u^1_{\rm rad}\sim100$ eV/cm$^3$ for $Z1$ for the starlight component, assuming a temperature of $T=4\,500$~K \citep[more details may be found in][]{VdeJ08b}. These become $u^0_{\rm rad}\sim1\,000$ eV/cm$^3$ and $u^1_{\rm rad}\sim40$ eV/cm$^3$ for Terzan~5 (note that the energy densities, and therefore the IC `bumps' associated with the upscattering of stellar photons, are linearly dependent on the assumed total stellar luminosity -- see Table~\ref{tab:andre}). For the CMB component, we use $u_{\rm CMB}\sim0.27$~eV/cm$^3$ (for $T=2.76$~K).

We next calculate the steady-state \textit{injection} electron spectrum for each zone by multiplying the cumulative ejection spectrum by an effective time-scale which incorporates both radiation losses and particle escape. We modeled the latter (i.e.\ trapping of particles by the nebular field) assuming Bohm diffusion. The injection spectra were calculated for nebular fields of $B=1\mu$G and $B=10\mu$G (BS07). Using these results, we obtained the expected unpulsed IC flux for each GC (see Figure~\ref{fig:Diff}). While the lower-energy SR spectra are not discussed here, we do include SR losses in the calculation of the steady-state injection particle spectrum \citep[see][]{VdeJ08b}.

\section{Discussion and Conclusions}
\label{sec:con}
\begin{figure}
\epsscale{0.7}
\plotone{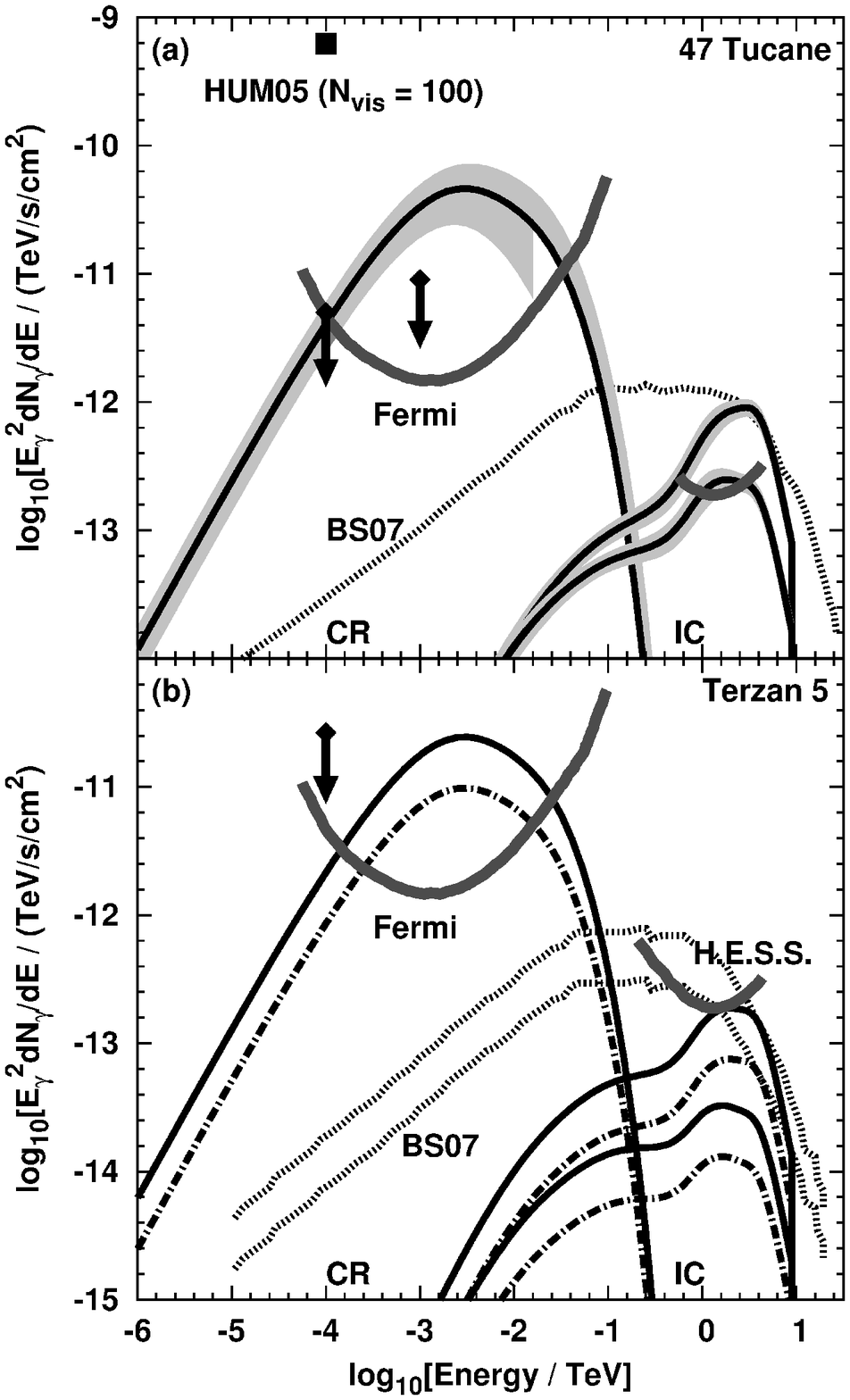}
\caption{Differential CR and IC spectra for 47~Tucanae (1a) and Terzan~5 (1b). The top IC spectra are for a nebular field $B=10\,\mu$G, while the bottom ones are for $B=1\,\mu$G. 
Dark gray lines indicate \textit{Fermi} \citep{Atwood09} and \textit{H.E.S.S.} \citep{Hinton04} sensitivities (appropriate for each GC). Arrows represent \textit{EGRET} upper limits. In (1a), we indicate a scaled prediction from HUM05, as well as $2\sigma$-bands (light gray bands) due to unknown pulsar geometry (the bottom CR one stops where $2\sigma$ exceeds the mean value). For both panels, we show predictions from BS07 (scaled to an average spin-down luminosity of $\dot{E}_{\rm rot}=2.2\times10^{34}$~erg\,s$^{-1}$). In (1b), solid lines are for $d=5.5$~kpc and dot-dashed ones for $d=8.7$~kpc ($2\sigma$-bands not shown).\label{fig:Diff}}
\end{figure}

We have modeled the HE and VHE components of CR and IC flux expected from a population of MSPs in the GCs 47~Tucanae and Terzan~5, for~100 members (Figure~\ref{fig:Diff}). For illustration, we assume that all MSPs are visible (i.e.\ $N_{\rm vis}=N_{\rm tot}=100$). However, the CR spectra are linearly dependent on $N_{\rm vis}$, while the IC spectra are linearly dependent on $N_{\rm tot}$.

\textit{Fermi} has recently released a list of bright sources \citep[$>10\sigma$,][]{Abdo09}, including preliminary fluxes of 47~Tucanae of $(5.1\pm2.0)\times10^{-8}$~cm$^{-2}$s$^{-1}$ and $(5.6\pm1.0)\times10^{-9}$~cm$^{-2}$s$^{-1}$ for energy bands spanning 100~MeV -- 1~GeV and 1~GeV -- 100~GeV respectively. Our average CR spectrum is overpredicting these fluxes by factors of $\sim2$ and~$\sim7$. However, the pulsed spectrum may quite reasonably be scaled to $N_{\rm vis}=50$ (as suggested by observations). Also, extrapolating from observed trends in single MSP spectra, one may speculate that it is possible that the cut-off energy may be smaller due to uncertainties in EOS parameters, the electric potential, magnetospheric structure, and pulsar geometry. For instance, taking $N_{\rm vis}=50$, and halving all CR spectral energies (i.e.\ shifting the CR spectrum to the left by a factor of~2), the model predictions improve to $\sim0.5$ and $\sim1.7$ times the \textit{Fermi} flux in these bands. 

Our ensemble-averaged CR spectrum, resulting from  $10^8$ single MSP spectra (i.e.\ $10^6$ 100-member cumulative spectra) with typical spectral cut-offs around a few GeV (generally depending on the individual MSP's $P$, $\dot{P}$, EOS, and geometry), exhibits a broad peak around $\sim3$~GeV. This average spectral cut-off energy follows directly from the assumed acceleration potential and B-field geometry of the PC model under consideration, as well as the fact that we simulate a random selection from the $P-\dot{P}$ parameter space for GC MSPs (similar to HUM05), and with different geometries. Given the fact that our CR spectrum seems to be reasonably successful at reproducing the preliminary \textit{Fermi} data, we expect that the predicted cut-off energy must be close to the real one. This implies that the average model acceleration potential must be of the right order of magnitude, assuming dipolar B-fields. A more detailed study involving further model and parameter space constraints will have to await further \textit{Fermi} spectral results.

We indicate \textit{EGRET} upper limits (ULs) from \citet{F95} on Figure~1a, and from \citet{Michelson94} on Figure~1b, converted assuming an $E^{-2}$ spectrum. Interestingly, the 100~MeV UL is close to the \textit{Fermi} sensitivity for 47~Tucanae. The \textit{EGRET} UL for Terzan~5 is not very constraining, and implies $N_{\rm vis}<363$ for $d=5.5$~kpc, and $N_{\rm vis}<907$ for $d=8.7$~kpc (using the average \textit{integral} CR spectrum).

As a reference, we have also included a CR prediction from HUM05, scaled to $N_{\rm vis}=100$, in Figure~\ref{fig:Diff}. HUM05 noted that the collective radiation from MSPs in GCs may be visible for \textit{Fermi}, and proceeded to apply their unscreened GR frame-dragging PC model to the MSPs in 47~Tucanae. They estimated the CR flux above 100~MeV from 47~Tucanae by integrating along a single magnetospheric B-field line, normalising to a (conservative) solid angle of 1~sr per PC, and scaling by the standard Goldreich-Julian PC current. This approach, which was followed to circumvent full 3D modeling, may be regarded as yielding a theoretical upper limit, given that different pulsar geometries (especially off-beam ones) have not expicitly been taken into account as done here, thus significantly exceeding our results at 100~MeV.

Predictions from BS07 focused on the possibility that HE leptons, which may be accelerated inside MSP magnetospheres and / or reaccelerated at colliding pulsar and stellar wind shocks, gradually diffuse through the GC and upscatter CMB and soft stellar photons to produce an unpulsed component in the GeV--TeV energy band. BS07 assumed that 1\% of the spin-down power (with the population average taken as $\dot{E}_{\rm rot}\sim1.2\times10^{35}$~erg\,s$^{-1}$) is injected in relativistic leptons. In the absence of detailed model predictions, they assumed power-law injection spectra (typical for shock acceleration) with various low and high-energy cut-offs and spectral indices, and normalized these using the assumed average MSP spin-down power and conversion efficiency. Representative resulting IC spectra, linearly scaled to our population-average spin-down luminosity of $\dot{E}_{\rm rot}=2.2\times10^{34}$~erg\,s$^{-1}$, are shown in Figure~\ref{fig:Diff}.
Our IC predictions, obtained by conservatively assuming no reacceleration of electrons, roughly agree with those of BS07 around a few TeV, but deviate at other energies due to different injection spectra, as well as slightly different assumptions regarding the energy density of stellar photons in the GCs (compare our two-zone approach with BS07's energy density profile given in their Figure~1).

\begin{figure}
\epsscale{0.7}
\plotone{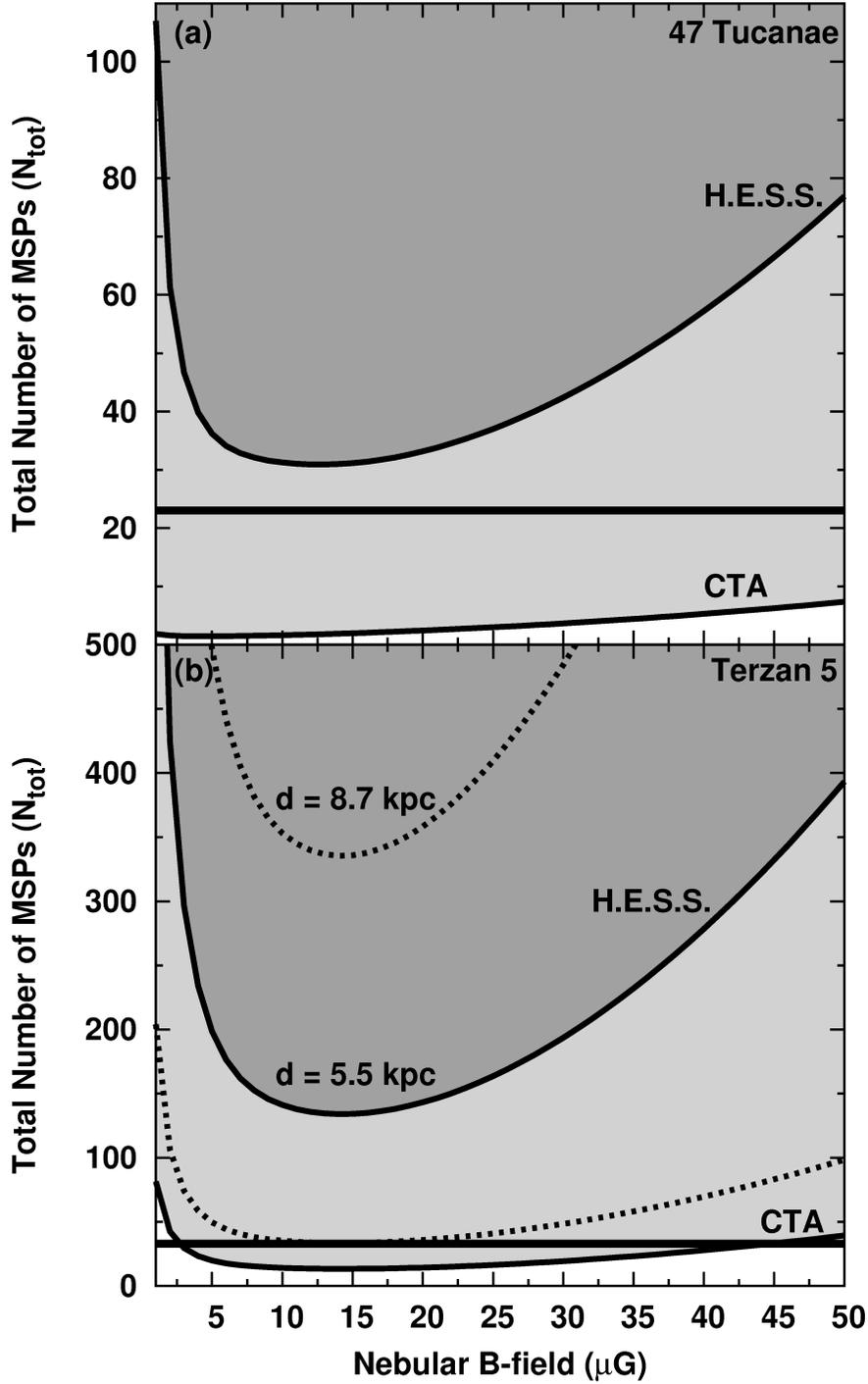}
\caption{Constraints on the total number of MSPs $N_{\rm tot}$ and nebular magnetic field $B$. Panel~(2a) is for 47~Tucanae, (2b) for Terzan~5. The dark gray indicate regions excluded by a \textit{H.E.S.S.} non-detection, while light (plus dark) gray regions are excluded by a \textit{CTA} non-detection (see text for details). In panel~(2b), the solid lines are for $d=5.5$~kpc, and dashed lines for $d=8.7$~kpc. The thick horizontal lines indicate the number of currently detected radio MSPs (see Table~\ref{tab:andre}).\label{fig:Constr}}
\end{figure}

Constraints on $N_{\rm tot}$ as well as $B$ may be derived from our model, using the predicted IC flux and assuming non-detection above 1~TeV at the level of 1\% and 0.1\% of the Crab flux, as achievable by \textit{H.E.S.S.} and \textit{CTA}, respectively. As shown in Figure~\ref{fig:Constr}, regions of the $B-N_{\rm tot}$ parameter space could be excluded from VHE ULs, e.g.\ choosing $B=10\,\mu$G for 47~Tucanae, the \textit{H.E.S.S.} UL constrains the total number of MSPs to $<33$ (and $<3$ for \textit{CTA}). Terzan~5 is less constrained (these limits become $<141$ and $<14$ for $d=5.5$~kpc, and $<353$ and $<35$ at $d=8.7$~kpc). 
For small $B$, particle escape is greater (and IC flux is lower), so that the constraints are less severe. This is also true for large $B$, where higher SR losses lead to inhibited IC flux. Furthermore, taking reacceleration of VHE electrons into account would increase SR and IC fluxes, implying more stringent constraints on $N_{\rm tot}$ and $B$ in the case of a non-detection. 

The $B-N_{\rm tot}$ constraints are however dependent on the choice of diffusion coefficient (a larger coefficient would result in enhanced particle escape and thus lower IC flux). It is thus possible to invert the argument and obtain constraints on the diffusion coefficient in the absence of reacceleration, using future VHE limits and assuming reasonable values for $N_{\rm tot}$ and $B$. To this end, a combination of \textit{Fermi} data and Monte Carlo modeling may help constrain $N_{\rm vis}$ and $N_{\rm vis}/N_{\rm tot}$, and thus $N_{\rm tot}$.
Also, measurements of the predicted diffuse ultraviolet SR \citep{VdeJ08b} and unpulsed diffuse gamma-ray IC spectra may further constrain the lepton injection spectrum as well as $B$. Lastly, if $B$ is large enough ($\ga20\,\mu$G) so that diffusion losses are unimportant relative to SR losses, particle reacceleration in the GC may be studied without having the concern of the latter being masked by energy-dependent diffusion.

\acknowledgments
We would like to thank Alice Harding, Dave Tompson, and Jarek Dyks for useful discussions. This research was supported by an appointment to the NASA Postdoctoral Program at the Goddard Space Flight Center, administered by Oak Ridge Associated Universities through a contract with NASA, and also by the South African National Research Foundation and the SA Centre for High Performance Computing.

\clearpage

%\clearpage

\end{document}